\documentclass[10pt,conference]{IEEEtran}
\IEEEoverridecommandlockouts
\usepackage{cite}
\usepackage{tikz}
\usepackage{amsmath,amssymb,amsfonts}
\usepackage{graphicx}
\usepackage{textcomp}
\usepackage{xcolor}
\usepackage{siunitx}
\usepackage{relsize}
\usepackage[caption=false,font=footnotesize]{subfig}
\usepackage{calc}
\usepackage{tabularx}

\usepackage{comment}
\usepackage{multirow}
\usepackage{array, hhline}
\def\BibTeX{{\rm B\kern-.05em{\sc i\kern-.025em b}\kern-.08em
    T\kern-.1667em\lower.7ex\hbox{E}\kern-.125emX}}

\usepackage{ragged2e}
\usepackage{booktabs}

\newcolumntype{L}[1]{>{\hsize=#1\linewidth\RaggedRight} X}

\begin{document}

\title{MEDA: Measurement-Efficient Disorder-Aware Majorana Zero Mode Detection in Realistic Devices}

\author{\IEEEauthorblockN{Nathan Jones\textsuperscript{*}, Binayyak Roy\textsuperscript{*}, Valentine Mohaugen, Ian Lewis, Toby Cox, Sumanta Tewari, and Rong Ge}
\IEEEauthorblockA{Clemson University, Clemson, South Carolina 29631\\
Email: \{najones, binayyr, valentm, iml, tscox, stewari, rge\}@clemson.edu}
\thanks{\textsuperscript{*}These authors contributed equally to this work.}
}

\maketitle
\begin{tikzpicture}[remember picture,overlay]
\node[anchor=south, yshift=15pt] at (current page.south) {
    \parbox{\dimexpr\textwidth-2cm\relax}{
        \centering \footnotesize \copyright~2026 IEEE. Personal use of this material is permitted. Permission from IEEE must be obtained for all other uses, in any current or future media, including reprinting/republishing this material for advertising or promotional purposes, creating new collective works, for resale or redistribution to servers or lists, or reuse of any copyrighted component of this work in other works.
    }
};
\end{tikzpicture}
\begin{abstract}
Fault-tolerant topological quantum computing relies on identifying Majorana zero modes (MZMs), but reliable detection in realistic devices remains challenging. Conventional topological indicators are inherently biased in finite, disordered systems, blurring the distinction between true MZMs and trivial states. Furthermore, attempts to map these indicators to real observables via machine learning require dense, expensive conductance measurements, creating a severe scaling bottleneck. To simultaneously address topological bias and measurement limitations, we present MEDA: a Measurement-Efficient, Disorder-Aware framework for MZM detection in realistic devices. MEDA maps sparse, practically obtainable observables directly to the robust periodic disorder invariant (PDI). Using a novel sparse parameter regime, MEDA reduces measurement volume by 10x while maintaining predictive quality, even in moderate to strong disorder regimes that limit conventional methods. Furthermore, MEDA naturally prioritizes input features consistent with the topological gap protocol, demonstrating strong physical interpretability. 
\end{abstract}


\section{Introduction}

Topological quantum computing offers a promising route to fault tolerance by encoding quantum information nonlocally in topological degrees of freedom, suppressing sensitivity to local noise and perturbations \cite{Nayak1996, Read2000, Moore1991, Nayak2008, Kitaev2003, wilczek1982quantum}. Majorana zero modes (MZMs), non-Abelian quasiparticles predicted to emerge in topological superconductors \cite{lutchyn2010majorana, oreg2010helical, sau2010non, sau2010generic}, are central to this paradigm because their nonlocal nature enables fault-tolerant operations through topological braiding. However, the practical realization of this paradigm critically hinges on reliable characterization of topological superconductivity in experimentally accessible, disordered devices.

Despite substantial experimental progress, robust MZM detection in realistic systems remains fundamentally challenging. Widely used topological invariants are typically derived under idealized assumptions. While successful in clean theoretical models, these tools become inherently biased \cite{Day2025} in finite, disordered devices, where impurities and spatial inhomogeneities blur the distinction between true topological MZMs and trivial near-zero-energy states \cite{mourik2012signatures, Deng2012, Das2012, rokhinson2012fractional, churchill2013superconductor, finck2013anomalous, deng2016majorana, zhang2017ballistic, chen2017experimental, nichele2017scaling, Mi2014, bagrets2012class, pikulin2012zero, prada2012transport, pan2020physical, moore2018two, Moore2018, vuik2019reproducing, stanescu2019robust}.

Prior work addresses this gap by adapting scattering matrix-based or local density of states (LDOS)-based topological indicators to finite-system settings \cite{stanescu2019robust,PhysRevB.111.104208,akhmerov2011,fulga2012,dassarma2016}. However, these  indicators still inherit boundary sensitivity and disorder-related biases. The recent periodic disorder invariant (PDI) offers a robust, bulk-defined topological indicator that remains well-defined in finite, disordered systems by construction \cite{PDI2025}. Unfortunately, these indicators require information that is unavailable outside of simulation, making them inaccessible for physical systems.

To overcome this lack of observability, recent works apply machine learning to infer topological indicators from experimentally measurable conductance data \cite{PhysRevLett.132.206602,CHENG20242507,PhysRevB.111.104208}. Such methods exhibit two key limitations. First, by predicting biased indicators, they inherit the same topological shortcomings which are often ineffectively addressed. Second, their input requires densely sampled conductance maps, which creates a severe measurement bottleneck in physical systems due to the inherently serial and painstakingly slow parameter-sweeping process \cite{gul2018ballistic, albrecht2016exponential,PhysRevB.107.245423,Baart_2016,Darulov__2020}. This data-acquisition bottleneck is exacerbated by complex topological indicators such as PDI, forcing existing methods to rely on the heavily biased predecessors. 

\begin{figure}[t]
    \centering
    \includegraphics[width=\linewidth]{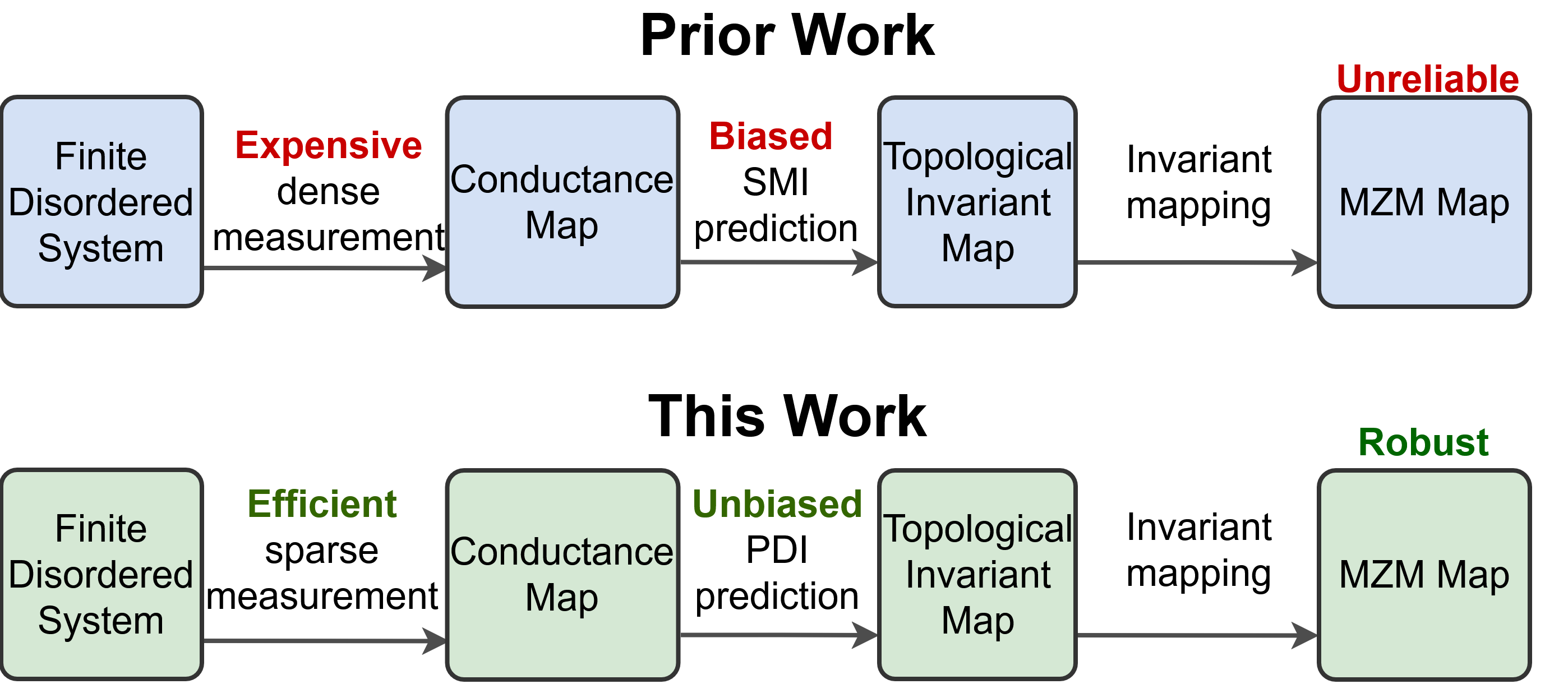}
    \caption{\textbf{MEDA Detection Pipeline.} Unlike traditional approaches requiring dense sampling and using biased  indicators, MEDA achieves scalable, bias-resilient MZM detection by mapping sparse conductance measurements directly to the disorder-aware PDI.}
    \label{fig:pipeline}
\end{figure}

In this work, we introduce \textbf{MEDA}, a \textbf{Measurement-Efficient, Disorder-Aware} framework designed to simultaneously address topological bias and experimental scalability, shown in \figurename \ref{fig:pipeline}. MEDA establishes the first direct mapping from experimentally measurable conductance data to the PDI, enabling  deployment of a bulk-defined, unbiased topological diagnostic in realistic experimental settings. Crucially, MEDA utilizes an attention-driven multiple instance learning (MIL) architecture to identify the most topologically informative regions of the conductance map. This enables MEDA's sparse-sampling paradigm, and consequently a $10\times$ reduction in required measurement volume with high classification accuracy. MEDA source code is available at https://doi.org/10.5281/zenodo.21478601.

We evaluate MEDA on a comprehensive dataset of simulated one-dimensional Majorana nanowire transport observables that explicitly targets regimes with moderate to strong disorder and fragmented phase diagrams. Our main contributions are as follows:

\begin{itemize}
    \item We propose MEDA, the first end-to-end pipeline mapping measurable conductance data directly to the binary PDI in finite, disordered semiconductor-superconductor (SM-SC) nanowires.

    \item We demonstrate that MEDA accurately reconstructs extended topological regions using only $10\%$ of the original measurement volume, resulting in massive savings in experimental acquisition time.

    \item We validate MEDA’s generalization under realistic device conditions with moderate-to-strong disorder, where conventional indicators often fail. 

    \item We show that the model's learned weighting independently prioritizes physically relevant conductance features consistent with established criteria, such as the topological gap protocol.
    
\end{itemize}

\section{Background and Motivation}
\label{sec:background}

While ideal one-dimensional SM-SC nanowires can host end-localized MZMs that offer fault-tolerant quantum computing \cite{Nayak2008}, realistic devices deviate significantly from idealized models, making MZM presence more ambiguous. Finite-size effects allow interaction between ideally independent end-states, while microscopic imperfections introduce disorder that generates trivial near-zero-energy states, such as partially separated Andreev bound states (ps-ABSs), that are difficult to distinguish from true MZMs despite offering no topological benefit  \cite{Kitaev2003,sau2010non,sau2010generic,lutchyn2010majorana,oreg2010helical, mourik2012signatures,Deng2012,Das2012,rokhinson2012fractional,churchill2013superconductor,finck2013anomalous,deng2016majorana,zhang2017ballistic,chen2017experimental,nichele2017scaling,albrecht2017transport,o2018hybridization,shen2018parity,sherman2017normal,vaitiekenas2018selective,albrecht2016exponential,Yu_2021,zhang2021,kells2012near}. These effects jointly produce false negatives and false positives under boundary-based diagnostics, including transport-based indicators \cite{akhmerov2011, PhysRevB.107.245423} and LDOS-based approaches \cite{PhysRevB.110.115436, Stanescu_2013,PhysRevLett.109.266402}.

The leading topological indicator for experimental systems, the scattering matrix invariant (SMI), attempts to formalize boundary-based diagnostics by characterizing topology through reflection properties of the system, ideally yielding a binary invariant \cite{fulga2012,Fulga_2011}. However, the SMI inherits the same fundamental limitations as other boundary-based indicators. Finite-size effects force continuous SMI values, and attempts to reduce false negatives via topological visibility (TV) \cite{pikulin2021protocolidentifytopologicalsuperconducting,PhysRevB.107.245423} amplify false positives from quasi-Majorana states \cite{Day2025,moore2018two,Moore2018,vuik2018reproducing}.

Additionally, SMI remains experimentally inaccessible despite its connection to transport. The full scattering matrix, from which SMI is derived, cannot be directly measured. Experimental proxies used to estimate the scattering matrix,  such as zero-bias conductance peaks, depend sensitively on fine tuning of external measurement conditions, rather than intrinsic system topology. As a result, the SMI remains biased in realistic regimes and impractical for experimental deployment.

\begin{figure}[t]
    \centering
    \includegraphics[width=\linewidth]{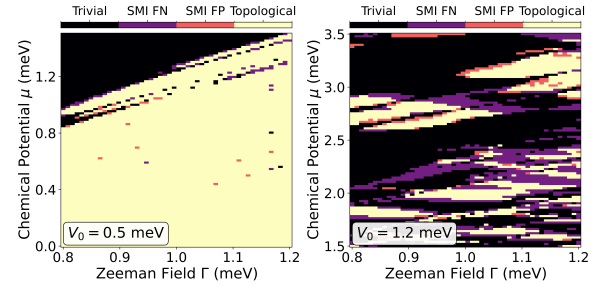}
    \caption{Compared to PDI, SMI performs well under weaker disorder (left), but produces significantly more false negatives (FN) and false positives (FP) under moderate disorder (right).}
    \label{fig:PDIvsSMI}
\end{figure}

In contrast, the PDI \cite{Day2025, PDI2025} provides a bulk-defined, unbiased diagnostic. Constructed by embedding the finite disordered nanowire into an infinite superlattice, the PDI yields a strictly quantized binary invariant that natively incorporates the full microscopic disorder profile. By eliminating boundary sensitivity, the PDI remains robust against finite-size effects, correcting SMI misclassifications, as shown in \figurename \ref{fig:PDIvsSMI}. Table ~\ref{tbl:smi_vs_pdi} summarizes PDI's advantages over SMI.

\begin{table}[t]
    \centering
    \caption{Due to its critical sensitivity to finite-size effects, disorder, and measurement tuning, the SMI cannot serve as a reliable diagnostic in real-world systems. The robust PDI must therefore serve as the ground truth, motivating our mapping approach.}
    \label{tbl:smi_vs_pdi}
    \begin{tabularx}{\linewidth}{L{0.18}| L{0.37} |L{0.33} } 
        \hline\hline
        \textbf{Property} & \textbf{SMI} & \textbf{PDI} \\
        \hline
        \textbf{Theoretical Basis} & Boundary scattering processes from determinant of scattering matrix $r$ & Bulk properties from system Hamiltonian \\
        \hline
        \textbf{Output Nature} & Continuous proxy on $-1 \leq \det(r) \leq 1$ & Strictly quantized and binary: $\{0, 1\}$ \\
        \hline
        \textbf{Observability} & Unobservable: matrix $r$ is unobtainable in practice & Unobservable: system Hamiltonian is unobtainable in practice\\
        \hline
        \textbf{Finite-Size Effects} & \textbf{Vulnerable:} susceptible to hybridization-induced false negatives  & \textbf{Robust:}  accounting for finite nanowire length \\
        \hline
        \textbf{Microscopic Disorder} & \textbf{Vulnerable:} yielding severe quasi-MZM-induced false positives & \textbf{Robust:} Accurately incorporating the full disorder profile \\
        \hline
        \textbf{Measurement Tuning} & \textbf{Vulnerable:} requiring fine-tuning of external lead coupling & \textbf{Robust:} independent of external measurement artifacts \\
        \hline\hline
    \end{tabularx}
\end{table}

While the PDI is theoretically robust, its reliance on the full system Hamiltonian prevents direct experimental deployment. Instead, experimentalists rely on boundary observables, such as differential conductance, measured over an extended parameter space that includes both control parameters and measurement variables. This creates a fundamental challenge: inferring bulk topology, and consequently the underlying topological phase diagram, from indirect, measurement-dependent observables.

Real-world measurement of these observables presents a severe bottleneck. Extracting  conductance signals in a dilution refrigerator is a slow and serial process that requires millisecond-scale integration time per measurement point \cite{PhysRevB.106.075306}. Sweeping the chemical potential ($\mu$) axis is particularly costly due to charge trap dynamics, cross-capacitance compensation, and electrostatic relaxation delays \cite{gul2018ballistic, albrecht2016exponential,PhysRevB.107.245423}. As a result, constructing a high-resolution phase diagram is prohibitively expensive \cite{Baart_2016,Darulov__2020}. While traditional methods concentrate samples within a narrow parameter window, potentially overlooking relevant regimes, MEDA achieves full parameter-space coverage without increasing the sampling budget by strategically subsampling conductance data, particularly along the costly $\mu$ axis.

\section{MEDA Framework}

\begin{figure}[t]
    \centering
    \includegraphics[width=\linewidth]{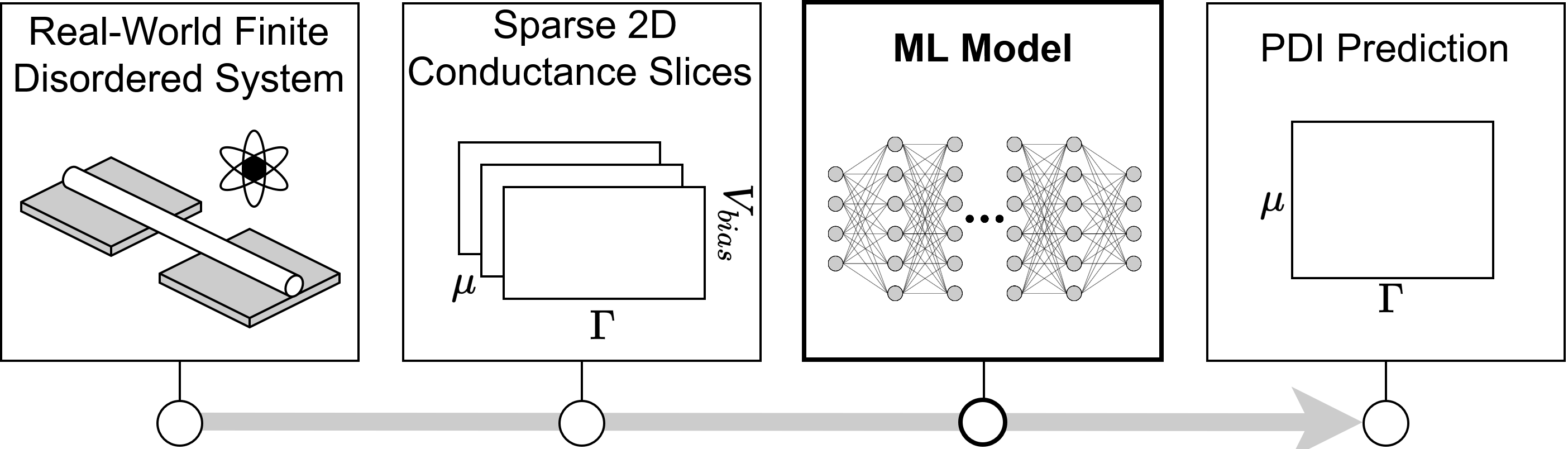}
    \caption{MEDA provides a scalable, bias-resilient pipeline for MZM detection for real-world devices by mapping sparsely sampled conductance measurements to the disorder-aware PDI.}
    \label{fig:overview}
\end{figure}

MEDA formulates the translation of sparse experimental observables into a bulk-defined topological invariant as an end-to-end supervised machine learning inference task, as shown in \figurename \ref{fig:overview}. MEDA's architecture targets three physics-induced structural challenges:
\begin{itemize}
    \item \textbf{Severe Class Imbalance:} Under realistic disorder, the topologically nontrivial phase space shrinks drastically, biasing standard models toward predicting fully trivial phase diagrams.
    \item \textbf{Phase Fragmentation:} High disorder shatters topological regions into disjointed islands. Unlike standard semantic segmentation tasks that identify highly continuous objects, this fragmentation challenges the network to balance  detection of localized phase boundaries against the risk of fitting to disorder-induced noise. 
    \item \textbf{Missing Data:} Reconstructing a full $\mu$-resolution phase map from sparse $\mu$ slices requires highly nonlinear interpolation across wide, unobserved chemical potential gaps.
\end{itemize}
The following section details MEDA's design considerations in light of these challenges. 

\subsection{Machine Learning Abstraction}

\textbf{Input Space ($\mathcal{X}$):} The input $X \in \mathcal{X}$ is a sparsely sampled ``bag" of $k$ discrete multi-channel differential conductance slices across the chemical potential $\mu$, denoted as $X = \{X_1, X_2, \dots, X_k\}$. To obtain $X$, we divide the full $\mu$ range into $100$ slices, and uniformly sample $k$ slices.  Each slice $X_i$ represents the $\Gamma, V_{bias}$ conductance profile for a single $\mu$ value. MEDA supports an arbitrary measurement budget $k$, deliberately subsampling along the $\mu$ axis by selecting $k\ll 100$ vastly expands the explorable parameter volume for a given measurement budget. Each slice $X_i \in \mathbb{R}^{N_{V_{bias}} \times N_\Gamma \times C}$ is a 3rd-order tensor representing a 2D spatial grid over the bias voltage $V_{bias}$ and the Zeeman field $\Gamma$, with a channel depth of $C=4$. The channel depth $C$ corresponds to the measurable local $(G_{LL}, G_{RR})$ and non-local $(G_{LR}, G_{RL})$ differential conductance channels. The subscripts on these channels refer to the left and right ends of the wire, with the first subscript denoting where voltage is applied to the system, and the second denoting where current is measured. The number of steps one dimension of our parameter grid are denoted $N_{V_{bias}}, N_{\Gamma}, N_{\mu}$ and can be calculated using table \ref{tbl:parameters}.

\textbf{Output Space ($\mathcal{Y}$):} The target output $Y \in \{0, 1\}^{N_\mu \times N_\Gamma}$ is a high-resolution, binary grid representing the system's ground-truth topological state, dictated by the PDI phase map over the full $(\mu, \Gamma)$ parameter space. While PDI is unaffected by $V_{bias}$, it changes sharply with $\mu$. Thus, we use the fully dense $\mu$ axis for the PDI map rather than the sparse $\mu$ input. A value of $1$ indicates a disorder-resilient, operationally topological MZM phase, while $0$ indicates a trivial state.

\textbf{Mapping and Objective ($f_\theta$):} We aim to learn a parameterized mapping $f_\theta$ that transforms the sparse input observables into a  probability phase map:
$$f_\theta: \mathbb{R}^{k \times N_{V_{bias}} \times N_\Gamma \times 4} \rightarrow [0, 1]^{N_\mu \times N_\Gamma}$$
where $\hat{Y} = f_\theta(X)$. Each element $\hat{Y}_{i,j}$ represents the model's predicted probability that the physical device exists in a topologically non-trivial state at the specific physical coordinates $(\mu_i, \Gamma_j)$. The neural network is optimized to minimize the discrepancy between the continuous prediction $\hat{Y}$ and the binary ground truth $Y$, yielding a surrogate model capable of inferring a global, bulk-defined topological invariant from partial surface observables without requiring full microscopic observability.

\subsubsection{Input Representation}
We abstract the multi-channel conductance slices into an image-like format, replacing standard RGB-Alpha channels with the four conductance channels. The spatial dimensions represent physical tuning parameters $V_{bias}$ and $\Gamma$. By enforcing sparsity along $\mu$, we treat the input as a ``bag" of semi-independent 2D images, allowing us to leverage Convolutional Neural Networks (CNNs).

CNNs are well-suited for this task due to their strong ability to learn spatial hierarchies and local correlations. Within our bag of 2D conductance images, experimental signatures of MZMs---such as zero-bias conductance peaks that remain stable over a finite range of $\Gamma$, or the closing and reopening of the bulk gap---manifest as continuous, visually distinct local features. The receptive field of a CNN naturally captures these contiguous physical phase boundaries.

Unlike standard optical images, however, the ``pixels" in our input possess strict, absolute physical units. Translation invariance, typically a core strength of CNNs, acts as a fundamental hurdle in this context: a conductance peak centered precisely at $V_{bias} = 0$ carries profound topological implications, whereas an identical peak translated to $V_{bias} = 0.05$ meV represents a physically distinct, likely trivial state. Furthermore, our input's discontinuity along the sparsely sampled $\mu$ axis requires specialized architectures capable of interpolating the missing information.

\subsubsection{Model Architecture and Complexity}
MEDA follows an encoder-decoder architecture to predict the phase map from conductance data. MEDA's encoding pipeline has two phases: independent single-$\mu$ analysis, and global $\mu$ interpolation and aggregation.

For the first task, MEDA employs a modified ResNet-18 shared encoder applied independently to each of the $k$ conductance slices. This provides sufficient depth to learn hierarchical transport features while remaining lightweight enough to prevent overfitting to localized microscopic disorder noise patterns. 

For the second task, MEDA analyzes the encoded slices in aggregate. To counteract CNN spatial invariance and provide necessary global context, we inject absolute coordinate values ($\mu, V_{bias}, \Gamma$) into the latent space via a learned embedding. Following feature extraction, the network utilizes a gated attention pooling mechanism. Because slices near phase transitions provide much more diagnostic utility than those deep in trivial regimes, this layer learns to apply weights each $\mu$ slice, filtering out uninformative regions and amplifying critical boundary signatures before aggregating them into a unified latent representation. 

Instead of outputting a single device-wide binary label like standard classifiers, MEDA utilizes a CNN decoder to project the aggregated latent representation into a high-resolution 2D spatial grid. Each pixel $P_{i,j}$ in the predicted phase diagram corresponds to a specific physical configuration $(\mu_i, \Gamma_j)$, with its magnitude representing the uncalibrated probability---or logit---of an operational MZM state. 

\subsection{Disorder-System Representation}

Realistic, disorder-diverse training data is essential for MEDA to map transport observables to the PDI robustly. In disordered nanowires, transport signatures depend on both controllable parameters and microscopic disorder realizations that are neither directly measurable nor reproducible across devices. To capture this intrinsic variability, we simulate finite nanowire systems over an extensive parameter space that includes both experimentally tunable parameters and latent disorder degrees of freedom, as summarized in Table~\ref{tbl:parameters}.

\begin{table*}[t]
\centering
\caption{Parameter space for this work. We choose ranges to be representative of real systems. Observability refers to whether a parameter is feasibly measurable/controllable in real systems.}
\label{tbl:parameters}
\small
\begin{tabularx}{\textwidth}{ |c|c|c|X|c|c| } 
  \hline
  \textbf{Parameter} & \textbf{Symbol} & \textbf{Observable} & \multicolumn{1}{|c|}{\textbf{Meaning}} & \textbf{Values} & \textbf{Granularity} \\
  \hline
  Disorder profile & $d$ & No & Array of point disorders corresponding to each nanowire lattice site & [0, 1] meV & N/A \\ 
  \hline
  Global disorder strength & $V_0$ & No & Root mean squared value of the disorder profile & [0.7, 2.5] meV & 0.18 meV \\ 
  \hline
  Correlation length & $l_c$ & No & Characteristic length scale over which disorder potentials exhibit spatial correlations & $20, 50, 90$ nm & N/A \\ 
  \hline
  \hline
  Chemical potential & $\mu$ & Yes & Chemical potential of system & $[1.5, 3.5]$ meV & $0.02$ meV\\
  \hline
  Bias voltage & $V_{bias}$ & Yes & Voltage difference across nanowire & $[-0.05, 0.05]$ meV & $3\times 10^{-3}$ meV\\
  \hline 
  Zeeman field & $\Gamma$ & Yes & Effective magnetic field that induces spin splitting in the nanowire & $[0.8, 1.2]$ meV & $8\times 10^{-3}$ meV\\
  \hline 
  \hline
  SM-SC coupling strength & $\gamma$ & Yes & Magnitude of superconductor influence on semiconductor nanowire & $0.2, 0.5$ meV & N/A \\ 
  \hline
  Dissipation & $\eta$ & Yes & Describes the leakage and transmission through the system & $0, 1\times 10^{-3}$ meV & N/A\\
  \hline 
  System length & $L$ & Yes & Length of the nanowire & $3$ $\mu$m & $10$ nm\\
  \hline
  Superconducting gap & $\Delta$ & Yes & Energy gap of the superconducting parent material & $0.3$ meV & N/A \\
  \hline
  RSO coupling strength & $\alpha$ & Yes & Coupling between electron spin and its motion in the nanowire & $140$ meV & N/A\\
  \hline
\end{tabularx}
\end{table*}

\textbf{Disorder Parameters:} Realistic modeling of microscopic disorder is central to our problem. Variations in disorder strength $V_0$, correlation length $l_c$, and spatial profile lead to substantial modifications of the low-energy spectrum. These variations fragment topological regions, shift phase boundaries, and crucially, produce trivial low-energy states whose nonlocal spectral weights closely masquerade as topological MZMs. To capture this intrinsic topological ambiguity, MEDA adopts a standard disorder model for SM nanowires \cite{Sau_2013,PhysRevB.110.115436,Stanescu_2013,PhysRevLett.109.266402}, parameterized by a site-specific profile $d\in [0,1]$, overall amplitude $V_0$, and correlation length $l_c$.

\textbf{Tunable System Parameters:} While the experimentally accessible parameters $\mu$, $\Gamma$, and $V_{bias}$ define the measurement space, disorder heavily distorts their relationship to topology. Consequently, transport measurements effectively act as projections of a higher-dimensional, heavily modulated parameter space. We deliberately select broad parameter ranges $(\mu \in [1.5,3.5]$~meV, $\Gamma\in[0.8,1.2]$~meV)\cite{PhysRevB.110.115436} to force MEDA to learn across regimes where disorder significantly shifts and fragments phase boundaries. MEDA's coverage of the $(\mu,\Gamma)$ parameter space ensures inclusion of regimes where disorder significantly alters phase structure causing conventional indicators to fail. This introduces a unique prediction challenge illustrated in \figurename \ref{fig:window}: unlike more conservative models like ViT~\cite{PhysRevB.111.104208} that become trivial under heavy disorder, MEDA is explicitly trained to navigate these heavily altered phases, positioning it as a viable tool for real-world MZM detection.

\textbf{Static System Parameters:} Variations in physical mechanisms—such as spectral broadening or Majorana hybridization—can alter the visibility of topological features in transport even when the underlying topology is unchanged. To isolate the effects of disorder while maintaining physically realistic device behavior, we fix static parameters to experimentally relevant values. These include Rashba spin-orbit coupling ($\alpha=140$~meV) \cite{stanescu2019robust,mourik2012signatures,Das2012,deng2016majorana}, weak and strong SM-SC coupling regimes ($\gamma=0.2,0.5$~meV),  $\eta=1\times10^{-3}$~meV to model quasiparticle broadening, and system length $L=3\,\mu$m to balance localization and finite-size effects. 

\subsection{High-Fidelity Training Data Synthesis}
\begin{figure}[t]
    \centering
    \includegraphics[width=0.9\linewidth]{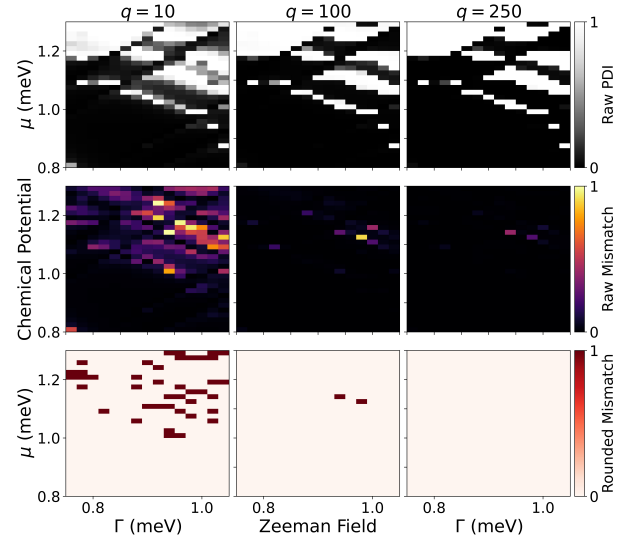}
    \caption{Top row: Reference raw PDI maps obtained at $q = 10,100,250$ iterations for disorder amplitude $V_0 = 1.2~\mathrm{meV}$ (left to right). Middle row: Difference between the raw PDI maps and the fully converged PDI map computed at $q=600$ (left to right). Bottom row: Corresponding differences between the rounded PDI value maps and the fully converged PDI maps.}
    \label{fig:Converge}
\end{figure}

Dense multidimensional conductance measurements are severely bottlenecked by experimental measurement time in real systems, but simulated systems are not so constrained. Therefore, we generate our dataset densely, then subsample the data during training. To generate our training dataset, we simulate realistic local and non-local differential conductance using the KWANT quantum transport package \cite{Groth_2014}. We include an optimal one-site potential barrier at the nanowire-lead interfaces to govern coupling strength, and a positive but finite dissipation term $\eta$ to capture quasiparticle leakage.

Ground-truth PDI labels are synthesized by targeting the disorder-aware PDI \cite{PDI2025}, which maps the device Hamiltonian to a 1D superlattice formalism to extract the bulk invariant. Because this evaluation is highly iterative, calculations are deployed across 191 CPU cores to generate dense phase maps. We enforce a rigorous convergence threshold defined in \figurename \ref{fig:Converge} to mitigate truncation errors and ensure a stable, binary topological ground truth for optimization.

\subsection{Physics-Guided Optimization}
\label{sec:optimization}

\subsubsection{Model-Driven $\mu$ Slice Selection}
As discussed in Section \ref{sec:background}, dense parameter sweeps are a major bottleneck for experimental studies of Majorana nanowires, especially along the $\mu$ axis. MEDA's incorporation of an expanded parameter space to capture shifting topological regimes exacerbates this measurement bottleneck. 

MEDA addresses this by subsampling along the $\mu$ axis and learning the underlying topological correlations to generate a latent representation covering the entire $\mu$ dimension. While our full simulated phase space spans $100$ adjacent $\mu$-slices, we use an attention-pooling mechanism which can extract bulk topological information by correlating non-trivial, fragmented conductance features across non-neighboring $\mu$ slices, enabling greatly reduced $\mu$ measurement budgets \cite{CHENG20242507, PhysRevLett.132.206602}. 

\subsubsection{Navigating the Topological Loss Landscape}
As disorder strength $V_0$ increases, the topological phase space drastically shrinks and shatters. Standard optimization strategies inevitably collapse in this landscape, falling into a trivial-majority local minimum where the model predicts a uniformly blank phase diagram. To prevent this collapse and force the MEDA to learn the physics of the phase boundaries, we construct a specialized, physically motivated composite loss function $\mathcal{L}$:
$$ \mathcal{L}_{total} = \lambda_1 \mathcal{L}_{BCE} + \lambda_2 \mathcal{L}_{Dice} + \lambda_3 \mathcal{L}_{Focal} + \lambda_4 \mathcal{L}_{TV} $$

Binary Cross-Entropy ($\mathcal{L}_{BCE}$) provides the foundational pixel-wise classification gradient, but it is insufficient for resolving fragmented topologies. We therefore integrate Focal Loss ($\mathcal{L}_{Focal}$) to dynamically down-weight the overwhelming trivial majority, forcing the gradient to focus on the rare, hard-to-predict pixels at quantum phase transitions. Dice Loss ($\mathcal{L}_{Dice}$)  evaluates spatial overlap, heavily penalizing the model when it misses small, fragmented MZM islands entirely. 

Finally, we introduce Total Variation ($\mathcal{L}_{TV}$) regularization to penalize overfragmentation. This smoothing is essential to prevent the model from hallucinating unrealistic noise artifacts, which we found more common when training on sparse inputs. However, this addition comes with a cost: MEDA prioritizes the identification of broad, operationally viable topological regions, resulting in limited ability to differentiate highly fragmented regions. 

The scalar weights ($\lambda_1 \dots \lambda_4$) act as model hyperparameters, and are optimized via hyperparamter grid search to stabilize the gradient and maintain hypersensitivity to subtle disorder-induced shifts. Trained models and weights are included in the released source code.

\section{Evaluation Methodology}
\subsection{Evaluation Metrics}
MEDA's performance is primarily evaluated using F1 score and precision. The parameter space suffers from severe class imbalance, as trivial phases dominate at higher disorder strengths, $V_0$. To prevent standard accuracy metrics from being skewed, we use the F1 score for a robust measure of global classification quality. Additionally, we heavily weight precision because false positives---mistaking a trivial state for a topological qubit---are highly detrimental to quantum computing reliability. An effective predictor must maintain both high precision and a strong global F1 score to be viable for real-world devices.

\subsection{Baseline Comparisons}
\begin{figure}[!t]
    \centering
    \subfloat[Topological region shift]{
        \includegraphics[width=\linewidth]{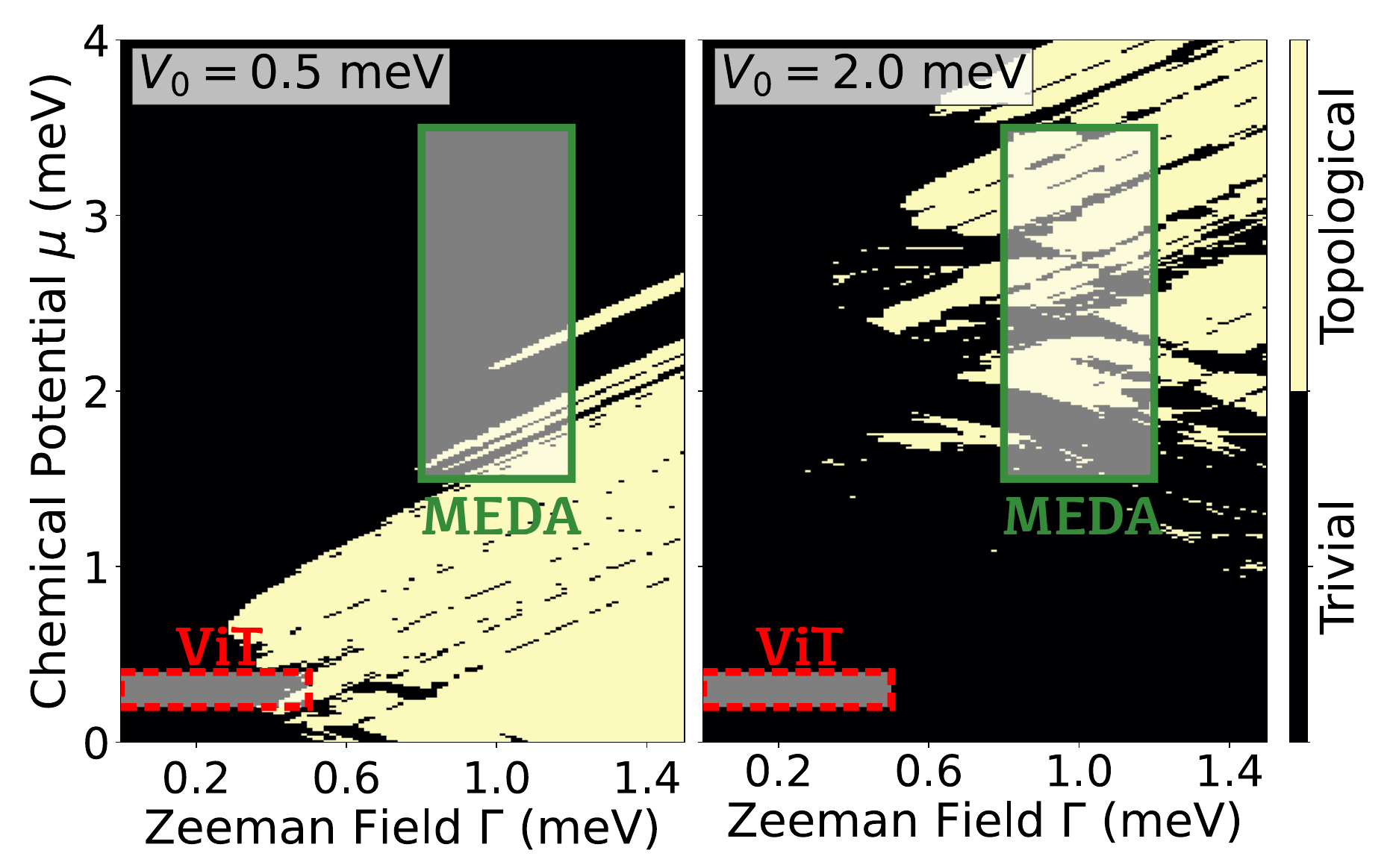}
        \label{fig:window}
    }
    \vspace{1em} 
    \subfloat[Topological activity]{
        \includegraphics[width=0.8\linewidth]{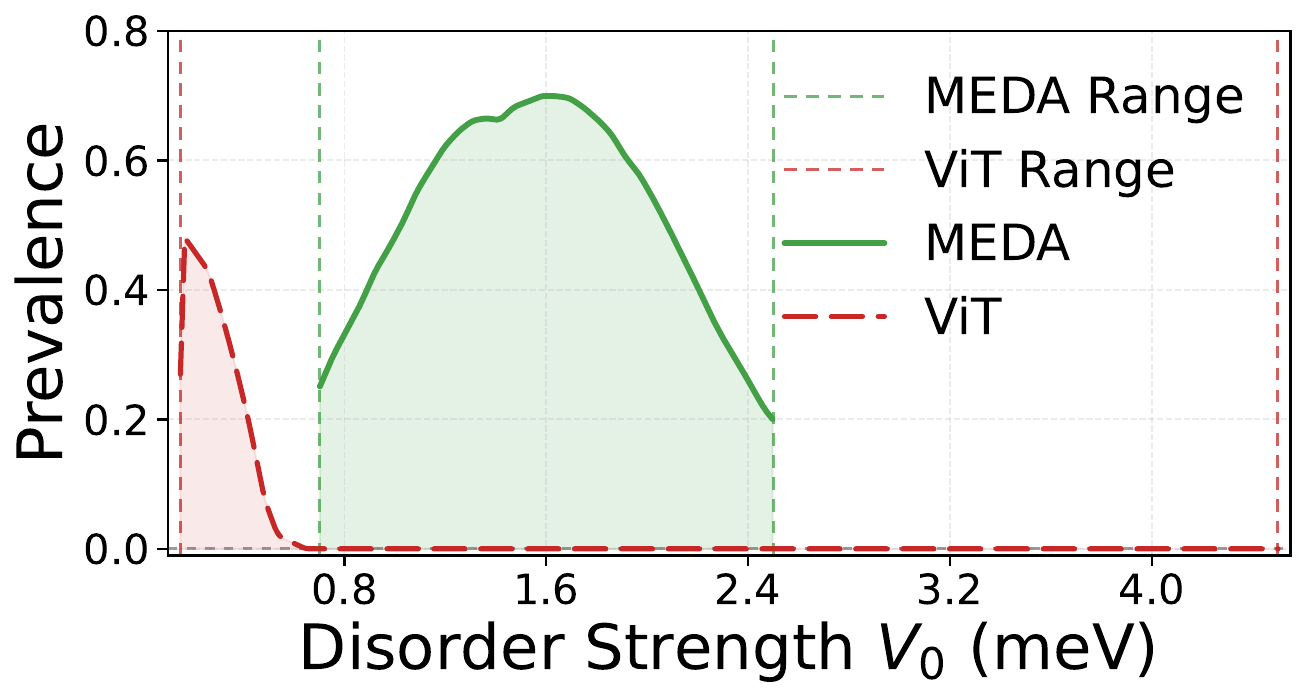}
        \label{fig:prevalence}
    }
    \caption{(a) As $V_0$ increases, MEDA's $\mu, \Gamma$ parameter window continues to capture topological activity, while ViT's window becomes trivial. (b) MEDA's disorder regime is optimized to capture topological activity, while ViT's regime is mostly trivial.}
    \label{fig:ParameterSpace}
\end{figure}

We compare MEDA against two counterparts: (1) the idealized theoretical SMI, which represents the  performance oracle and upper limit of SMI-based detection, assuming full knowledge of the system.  (2)  the  ViT framework \cite{PhysRevB.111.104208}, which is the state-of-the-art learning-based MZM detection using measurements. All comparisons are evaluated relative to the ground-truth PDI. Because  the ViT model is  closed-source, not available for duplication, we directly cite  the results reported in~\cite{PhysRevB.111.104208}, focusing on their full disorder regime spanning $V_0 \in [0.15, 4.5]~\mathrm{meV}$. Since ViT predicts a biased SMI map rather than a bulk invariant, we use Bayes' rule to incorporate probabilistic error from SMI biases. As ViT outputs continuous SMI values, we compute the F1 score using a threshold of $\mathrm{SMI}=0$, following~\cite{PhysRevB.111.104208}.

We note that ViT is trained over a smaller and coarser parameter space, leading to a predominance of trivial SMI maps, particularly at higher disorder strengths. In contrast, MEDA operates over a broader and more finely resolved parameter space, ensuring substantial coverage of regimes containing topological phases, as illustrated in Figs.~\ref{fig:window} and \ref{fig:prevalence}. While ViT covers a larger disorder regime than MEDA, Fig. \ref{fig:prevalence} demonstrates that this range is mostly trivial for ViT's parameter window in Fig. \ref{fig:window}. Crucially, while MEDA targets a smaller disorder regime, the entire regime is topologically active, resulting in a larger topologically-nontrivial regime than ViT's.

Furthermore, ViT is trained over a range of correlation lengths $l_c \in [20,70]$~nm, with $l_c$ randomly sampled for each disorder realization. As correlation length information is not explicitly reported, we compare against models trained at a fixed $l_c = 50$~nm, acknowledging the resulting mismatch. While this training mismatch inherently blurs a purely direct architectural performance delta between MEDA and ViT, we emphasize that MEDA’s capacity to learn across a substantially broader and topologically active parameter space establishes it as a more comprehensive tool for real-world phase mapping. Given this training distribution, ViT's predictive performance is expected to degrade at smaller correlation lengths (e.g., $l_c = 20$~nm), where disorder dominates.

\subsection{Evaluation Dataset}
The evaluation dataset is constructed using the same parameter space defined in Table~\ref{tbl:parameters}, with unseen disorder profiles not included in the training set. Strict separation between training and evaluation disorder profiles is necessary to ensure MEDA generalizes to real devices, which exhibit unique microscopic disorder profiles that cannot be reproduced or controlled. This setup reflects the intrinsic variability of experimental systems and tests MEDA's ability to learn disorder-robust features rather than memorizing specific realizations.

\subsection{Evaluation Studies}
Evaluating MEDA requires quantifying not only predictive accuracy, but also its performance under the intrinsic constraints of disordered, measurement-limited systems. To this end, the evaluation framework is partitioned into four complementary analyses, each probing a distinct aspect of the underlying inverse problem: accuracy relative to theoretical limits, robustness under sparse measurements, interpretability of learned features, and stability across disorder regimes.

\subsubsection{Global Pipeline Efficacy}
We first establish baseline predictive performance relative to both theoretical limits and existing methods. MEDA is evaluated using the global F1 score across 100 unseen disorder profiles at 10 different disorder amplitudes $V_0$. We compare against the two baselines, quantifying how closely MEDA approaches the SMI limit while improving upon state-of-the-art, thereby establishing MEDA as both theoretically effective and practically competitive.

\subsubsection{Resolution and Throughput Trade-offs}
A central challenge in experimental deployment is the trade-off between measurement sparsity and predictive accuracy. To characterize this, we systematically vary the fraction of available input data during inference. This analysis quantifies the efficiency with which MEDA extracts topological information from incomplete measurements and provides a direct calibration of how predictive accuracy varies as a function of measurement sparsity providing a direct estimate of the performance achievable with a given fraction of the full dataset.

\subsubsection{Mechanistic Interpretability via Feature Extraction}
Given the disorder-induced ambiguity between transport signatures and topology, it is essential to verify that MEDA learns physically meaningful features rather than spurious correlations. To probe this interpretability, we analyze the model’s attention weights. By extracting the conductance maps in $(\Gamma, V_{bias})$ space that receive the highest attention scores, we can directly compare the model's learned focus against established MZM-indicative transport features, such as those defined by the topological gap protocol (TGP) \cite{PhysRevB.107.245423}. This methodology allows us to verify whether MEDA's predictions are grounded in physical reality without providing the model any explicit prior knowledge of these protocols.

\subsubsection{Robustness in Disordered Regimes}
MEDA's robustness against disorder is of significant concern for practical deployment, as guaranteed  microscopic inhomogeneities can significantly alter transport signatures without changing the underlying topology. We evaluate MEDA across 100 distinct disorder realizations spanning low, medium, and high correlation regimes $l_c \in \{20, 50, 90\}$~nm, while varying  disorder strength over $V_0 \in [0.75, 2.5]$~meV. 
This analysis maps the operational boundaries of MEDA in the presence of disorder, a critical requirement for translating model predictions into experimentally actionable confidence levels.
\section{Results}
\label{sec:results}
\subsection{Global Pipeline Quality}
\begin{figure}[t]
    \centering
    \includegraphics[width=0.8\linewidth]{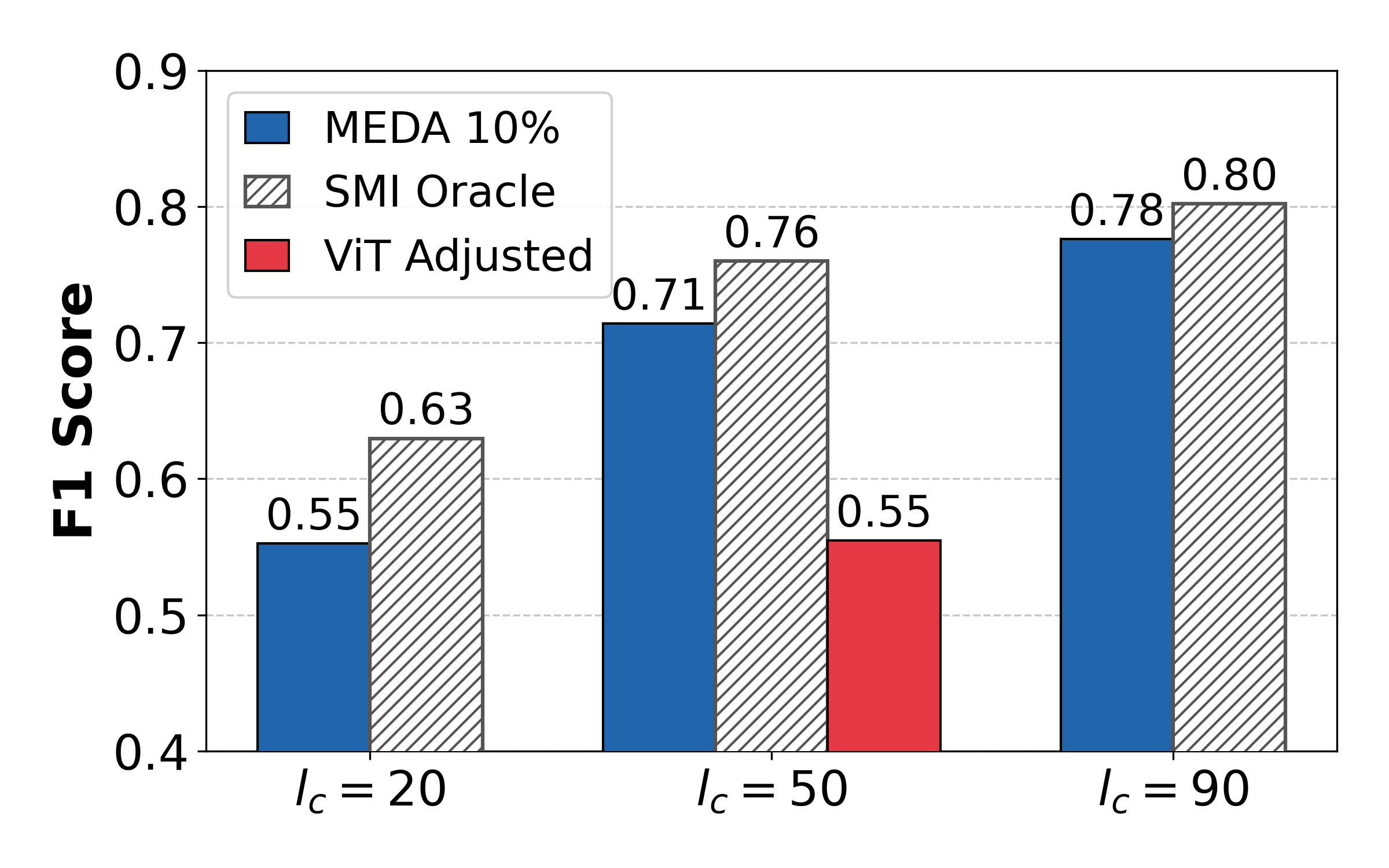}
    \caption{Despite using only $10\%$ of observable data, MEDA achieves performance comparable to the SMI Oracle and offers significant improvement over ViT, the state-of-the-art SMI-based prediction pipeline.}
    \label{fig:performance_overview}
\end{figure}

MEDA achieves high global predictive quality while operating with only $10\%$ of available data, demonstrating its viability as a measurement-efficient MZM detection pipeline. As shown in \figurename~\ref{fig:performance_overview}, MEDA attains performance comparable to the SMI Oracle—the theoretical ceiling for SMI-based approaches—while outperforming ViT \cite{PhysRevB.111.104208}, the current state-of-the-art SMI-based pipeline, despite targeting a more complex parameter space.

MEDA’s performance exhibits a clear dependence on disorder correlation length, reflecting its disorder-aware design. The gap between MEDA and the SMI Oracle is largest at low $l_c=20$, where highly jagged disorder profiles produce less predictable transport signatures. As $l_c$ increases, smoother disorder leads to more stable topological features and correspondingly improved predictive accuracy.

\subsection{Resolution and Throughput Tradeoffs}
\begin{figure}[t]
    \centering
    \includegraphics[width=0.7\linewidth]{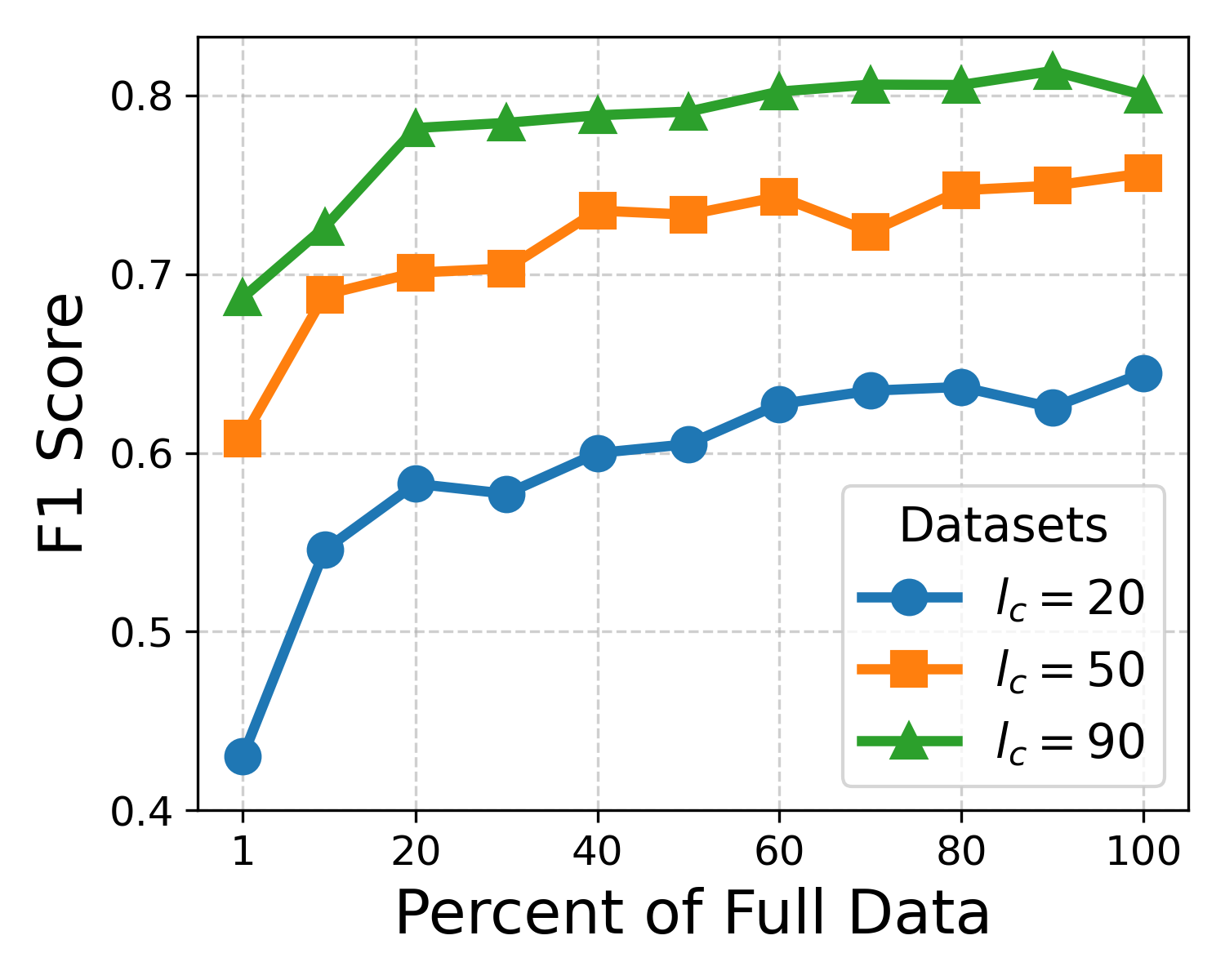}
    \caption{F1 Score vs Data Usage, with training limited to 500 epochs. Notice that past $10\%$ of full data, increasing data has diminishing returns.}
    \label{fig:tradeoff}
\end{figure}

MEDA enables a tunable trade-off between measurement cost and predictive accuracy, directly addressing experimental data acquisition constraints. \figurename~\ref{fig:tradeoff} shows a logarithmic relationship between the fraction of input data used during inference and the resulting F1 score. While MEDA operates at $10\%$ input data to significantly reduce measurement burden, the framework allows flexible adjustment depending on experimental requirements.

The performance curve reveals diminishing returns beyond modest data fractions, validating MEDA’s measurement-efficient design. In particular, systems with $l_c=90$ maintain high F1 scores with as little as $1\%$ of full conductance data, indicating that accurate topological inference can be achieved with extremely sparse measurements. This suggests the potential for near real-time analysis of MZM systems under favorable disorder conditions.

For subsequent evaluations, we fix the input fraction at $10\%$ to reflect a practically achievable measurement regime. As we will demonstrate in our throughput analysis, this specific fraction provides an optimal trade-off, maximizing experimental viability while preventing the diminishing returns observed at denser sampling rates. This choice prioritizes practicality for real-world systems while maintaining strong predictive performance.

\subsection{Physics-Informed Feature Extraction}
\begin{figure*}[t]
    \centering
    \includegraphics[width=0.8\linewidth]{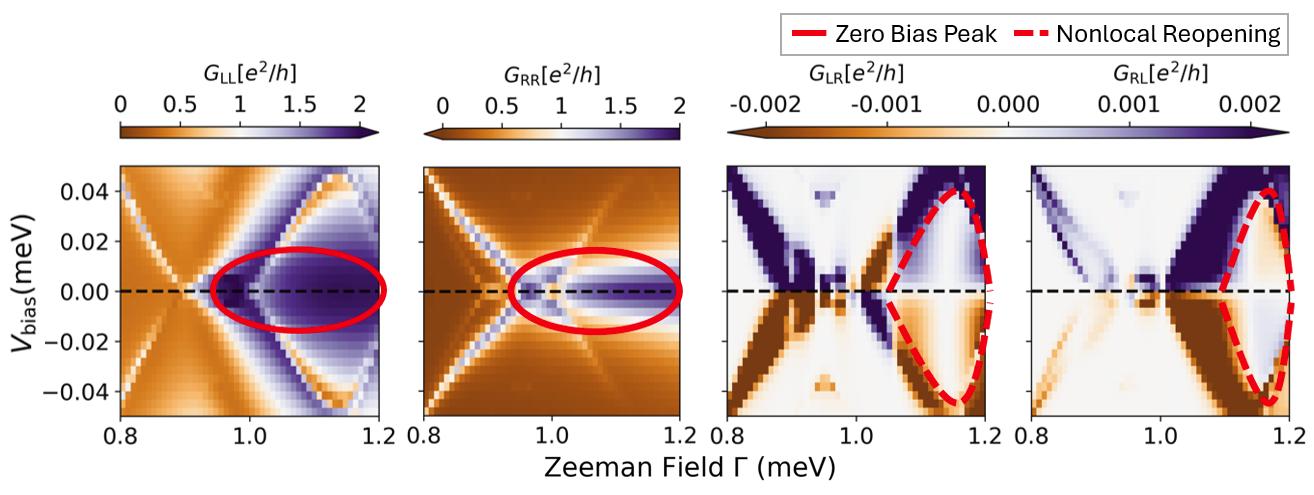}
    \caption{MEDA's trained attention weights heavily prioritize conductance maps with clearly visible conductance features that Morteza et al. establish as indicators of high topological phase probability \cite{PhysRevB.107.245423}. We circle sustained zero bias peaks (ZBP) in local conductance plots  $G_{LL},G_{RR}$, and indicate the conductance gap reopening in the nonlocal conductance plots $G_{LR},G_{RL}$. For the $\Gamma > 1.0$ region, the simultaneous ZBP and reopened nonlocal conductance gap are indicators of high topological phase probability. As far as we are aware, MEDA is the first machine learning approach to demonstrably learn physical MZM-heralding features without guidance specifying those features.}
    \label{fig:tgp}
\end{figure*}

MEDA learns physically meaningful features by selectively prioritizing informative chemical potential slices, demonstrating that the model identifies topological signatures directly from conductance data rather than relying on spurious correlations. \figurename~\ref{fig:tgp} shows an illustrative example of a single $\mu$-slice conductance profile set, selected from the conductance slices with the highest $0.01\%$ of attention scores across the entire evaluation dataset. 
 
The selected high-attention conductance profiles exhibit clear signatures consistent with the topological gap protocol (TGP) \cite{PhysRevB.107.245423}, including zero-bias peaks in local conductance channels and gap closing and reopening features in non-local conductance channels. Importantly, this behavior emerges without any explicit encoding of TGP-based criteria in the model, indicating that MEDA autonomously learns to identify physically meaningful topological features. This constitutes, to our knowledge, the first instance of a model demonstrating the ability to prioritize embedded topological information aligned with established experimental protocols without any protocol-specific training.

These results demonstrate that MEDA’s feature selection mechanism is grounded in physically interpretable transport signatures, enabling targeted identification of high-information regions in parameter space. By demonstrating this capability, MEDA provides a principled foundation for designing measurement strategies that prioritize informative $\mu$ slices over uniform sampling.

\subsection{Robustness in Disordered Regimes}

MEDA maintains strong predictive performance across varying disorder strengths, establishing its reliability in realistic nanowire environments. While performance degrades with increasing disorder strength $V_0$, MEDA consistently remains competitive as a practical MZM detection method. We note that all results in this section are obtained under a fixed measurement constraint of $10\%$ input data. As demonstrated in \figurename~\ref{fig:tradeoff}, increasing the available data fraction leads to systematic improvements in predictive accuracy, indicating that the observed degradation at higher $V_0$ and lower $l_c$ primarily reflects measurement sparsity rather than a fundamental limitation of the framework.

\subsubsection{Low Disorder-Correlation Regime $l_c=20nm$}
\begin{figure}[t]
    \centering
    \includegraphics[width=0.85\linewidth]{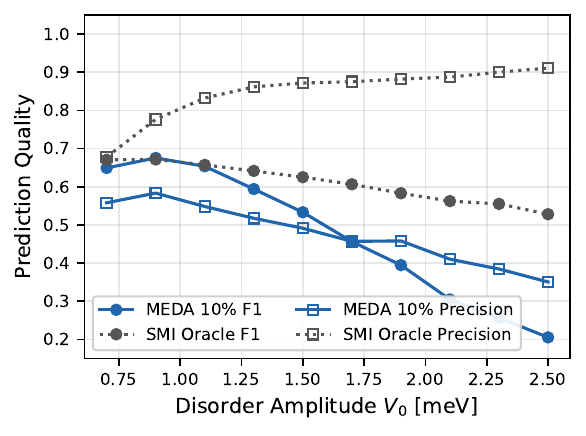}
    \caption{In the low disorder-correlation regime ($l_c=20nm)$, MEDA performs well for lower $V_0$ disorder, but struggles at higher $V_0$ due to significant fragmentation and more-pronounced class imbalance.}
    \label{fig:l20_f1}
\end{figure}

The low disorder-correlation regime represents the most challenging setting for disorder-aware prediction due to highly jagged disorder profiles. As shown in \figurename~\ref{fig:l20_f1}, MEDA achieves lower baseline F1 scores in this regime, with performance degrading more rapidly as $V_0$ increases. Nevertheless, MEDA retains useful predictive capability into moderate disorder strengths.

SMI also shows reduced effectiveness in this regime, reflecting its lack of disorder awareness. Jagged disorder profiles amplify discrepancies between disorder-aware and disorder-unaware diagnostics, limiting SMI’s reliability.

ViT is expected to show limited sensitivity to increasing disorder in this regime due to its restricted parameter window. Because its predictions collapse into predominantly trivial classifications at higher disorder, its F1 score flattens out, reflecting a complete loss of predictive sensitivity rather than robustness. If evaluated over the broader parameter space used by MEDA, a similar degradation trend would likely emerge.

\subsubsection{Medium Disorder-Correlation Regime $l_c=50nm$}
\begin{figure}[t]
    \centering
    \includegraphics[width=0.85\linewidth]{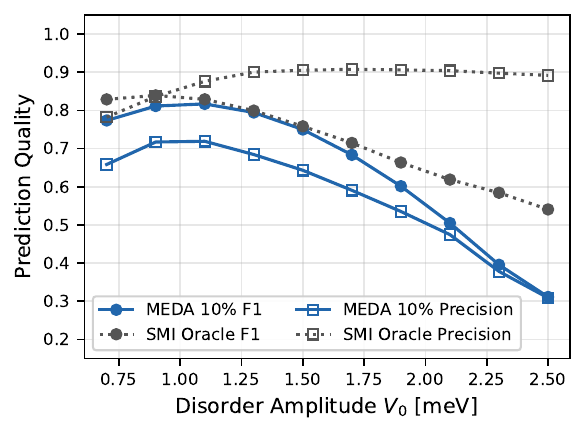}
    \caption{MEDA maintains predictive performance comparable to the theoretical SMI baseline up until high disorder ($V_0$) levels in the medium disorder-correlation regime ($l_c=50\text{nm}$), highlighting its viability as an indicator for real-world devices.}
    \label{fig:lc50_f1}
\end{figure}

In the medium disorder-correlation regime, MEDA achieves performance comparable to SMI across a broad range of disorder strengths, demonstrating its effectiveness in experimentally relevant conditions. While SMI serves as a useful reference, its reliance on inaccessible full-system information limits its direct applicability. In contrast, MEDA’s comparable performance, combined with its reliance on experimentally accessible inputs, positions it as a practical methodology for real-world MZM detection. MEDA is particularly well positioned compared to widely-used SMI-based pipelines, which compound prediction errors with errors inherent in their SMI target.

\subsubsection{High Disorder-Correlation Regime $l_c=90nm$}
\begin{figure}[t]
    \centering
    \includegraphics[width=0.85\linewidth]{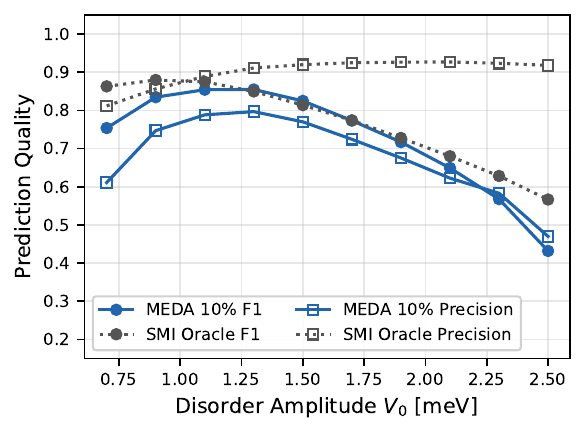}
    \caption{In the high disorder-correlation regime ($l_c=90\text{nm}$) smoother disorder profiles allow MEDA to retain high predictive accuracy even at elevated disorder strengths.}
    \label{fig:lc90_f1}
\end{figure}

MEDA achieves its strongest performance in high disorder-correlation regimes, where smoother disorder profiles preserve topological structure. As shown in \figurename~\ref{fig:lc90_f1}, MEDA remains more stable with increasing disorder strength $V_0$ compared to lower $l_c$ regimes. The improved performance reflects the increased stability of transport signatures under smoother disorder, allowing MEDA’s feature extraction to operate with higher confidence. This trend confirms that disorder correlation length is a key factor governing predictive reliability.

Importantly, MEDA can exceed SMI performance in certain low-disorder regimes within this setting. Because the SMI is fundamentally boundary-focused and susceptible to quasi-Majorana false positives, MEDA's disorder-aware, bulk-oriented mapping provides a more accurate reflection of the true PDI ground truth. This result highlights the limitations of relying on boundary-based invariants in realistic systems.

\subsection{Qualitative Analysis}
\begin{figure}[t]
    \centering
    \includegraphics[width=0.9\linewidth]{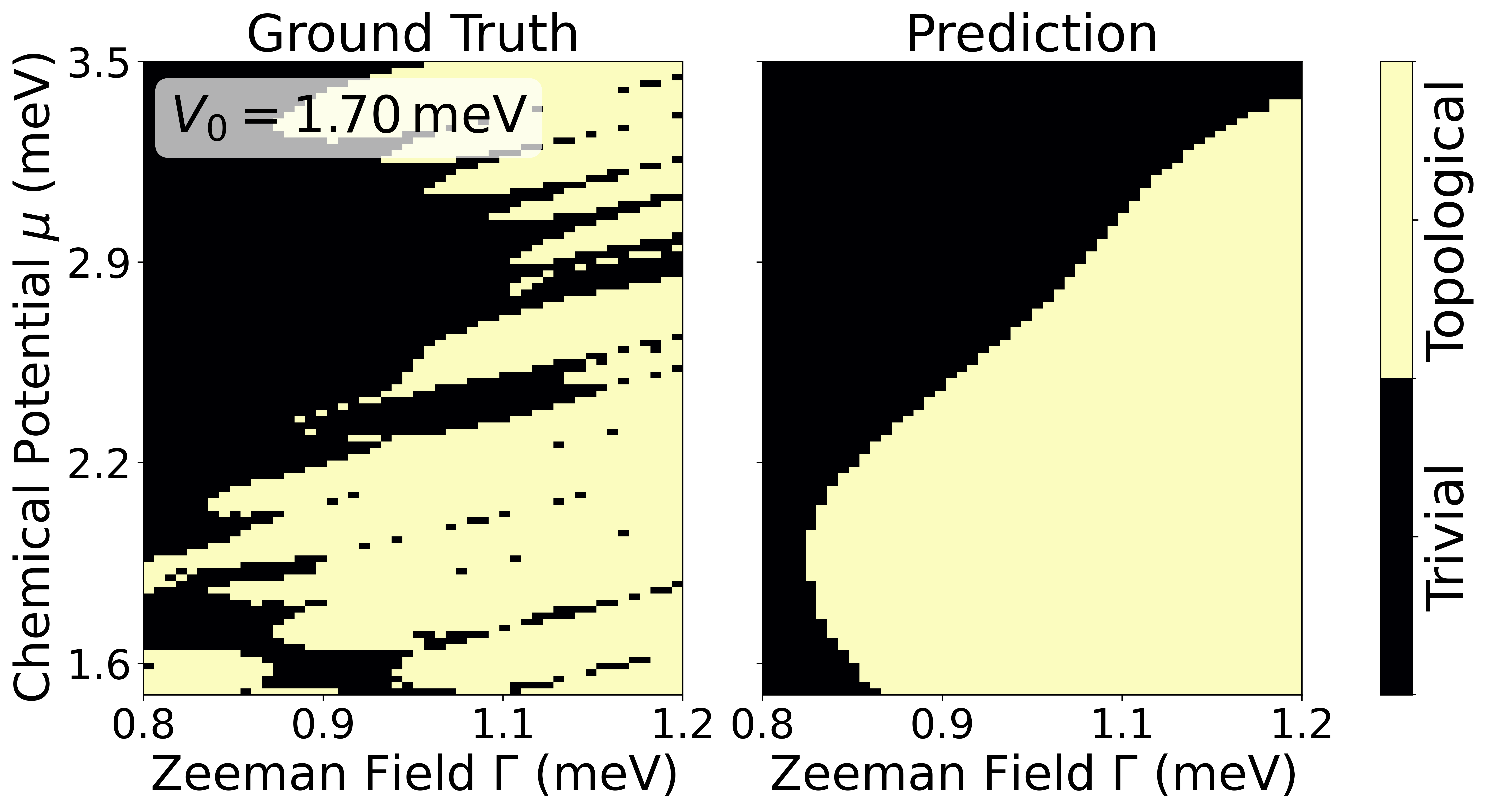}
    \caption{MEDA predicts smooth topological phase transitions and continuous regions. While this helps to avoid overfitting during training, it leads to misclassifications for heavily fragmented topological phases under high disorder.}
    \label{fig:prediction}
\end{figure}

While MEDA exhibits strong predictive performance relative to existing work, it remains challenged by heavy fragmentation and chaotic topological phase boundaries. \figurename \ref{fig:prediction} shows that MEDA over-prioritizes smooth phase boundaries and consistent regions, resulting in erroneously simplified predictions of highly fragmented landscapes. 

We attribute MEDA's tendency to oversimplify landscapes to our extensive optimization against overfitting. Because each real-world system is guaranteed a unique disorder profile, it is essential to avoid overfitting to the training dataset. Many overfit-prevention techniques---including MEDA's training input noise injection and total variation regularization---actively encourage smoother prediction maps that may not accurately represent the highly-fragmented topological phase maps at high disorder. Significant analysis is required to uncover the optimal balance between overfit prevention and fragmentation prediction; we leave this to future work. 
\section{Related Work}
MEDA is an end-to-end framework designed to predict MZMs in real-world, 1D nanowire devices. We situate MEDA's contributions relative to three prevailing paradigms of MZM detection: idealized theoretical indicators, hardware-driven experimental diagnostics, and recent data-driven machine learning frameworks.

Foundational theoretical models establish the basis for MZMs through bulk invariants and gap-closing signatures\cite{Kitaev2003,sau2010non,sau2010generic,lutchyn2010majorana,oreg2010helical}. However, as detailed in Section \ref{sec:background}, attempting to adapt these concepts into finite-system indicators---such as the SMI \cite{fulga2012,Fulga_2011,Day2025,pikulin2021protocolidentifytopologicalsuperconducting} or local density of states \cite{Stanescu_2013,PhysRevLett.109.266402,PhysRevB.110.115436}---results in boundary-induced biases and severe false positives in the presence of disorder. On the experimental front, hardware validation relies heavily on tunneling spectroscopy \cite{mourik2012signatures,Deng2012,Das2012,deng2016majorana,zhang2017ballistic,albrecht2016exponential,akhmerov2011} and advanced correlation methods like the TGP \cite{ZhangNature2018, prada2012transport}. While these experimental approaches provide and validate necessary transport observables, they face severe measurement scalability bottlenecks and interpretability challenges when attempting to construct global phase diagrams.

Recent literature has increasingly turned to machine learning to connect theoretical indicators and experimental observables. Cheng et al. demonstrated that ML models can successfully extract disorder-dependent topological information from conductance data \cite{CHENG20242507}. Similarly, Taylor and Sarma deployed a Vision Transformer architecture to reproduce SMI directly from conductance maps \cite{PhysRevB.111.104208}. Other data-driven frameworks aim to infer underlying disorder characteristics from transport measurements \cite{PhysRevLett.132.206602}.

While these ML approaches represent a significant step forward, they remain fundamentally limited by two constraints: topological bias and measurement scalability. Because existing models predominantly target the SMI, their predictions inherit its boundary-related biases, resulting in the incorrect classification of quasi-Majorana states and phases obscured by finite-size effects. Furthermore, these architectures rely on dense conductance maps, failing to account for the prohibitively high serial measurement cost of sweeping the chemical potential in laboratory settings.

MEDA distinguishes itself from other works relating observables to invariants in two critical ways. 
\begin{enumerate}
    \item \textbf{Targeting an Unbiased Ground Truth:} To our knowledge, MEDA is the first ML framework that eliminates boundary-induces biases by targeting the bulk-defined, disorder-aware PDI \cite{PDI2025}, providing resiliency against quasi-Majorana states and finite-size effects.
    \item \textbf{Measurement-Efficient Inference:} Rather than requiring dense parameter sweeps, MEDA utilizes deliberately sparse conductance data. By explicitly learning to prioritize high-information chemical potential slices—--autonomously extracting features consistent with established experimental criteria like the TGP—--MEDA provides a practical pathway for identifying topological regimes with a $10\times$ reduction in experimental overhead.
\end{enumerate}

\section{Conclusion}
In this work, we present MEDA, a measurement-efficient framework for the robust detection of MZMs in finite, disordered systems. By directly mapping experimentally accessible transport observables to the disorder-aware PDI, MEDA successfully avoids the boundary-induced biases and false positives inherent to conventional indicators like the SMI.

Crucially, MEDA overcomes the severe serial data acquisition bottleneck that currently limits experimental scalability. Using an attention-based architecture to aggregate sparse conductance measurements, the model achieves up to a $10\times$ reduction in required input data with minimal loss in predictive accuracy. MEDA remains robust even under strong disorder and finite-size effects where traditional indicators fail, providing a highly scalable and practical pathway for mapping topological phase boundaries in realistic devices.

While MEDA represents a significant advance in scalable MZM detection, certain limitations remain. Specifically, the strong regularization required to prevent overfitting to unique disorder profiles currently leads the model to oversimplify highly fragmented topological landscapes at extreme disorder strengths. Future works must determine a more optimal balance between noise suppression and high-resolution phase boundary prediction. Potential methodologies may include adaptive regularization schemes that dynamically scale the $\mathcal{L}_{TV}$ penalty in response to high local variance, or alternative loss functions that penalize phase boundary mismatches without aggressively smoothing localized topological islands. Moving forward, the physical interpretability of MEDA’s attention mechanism offers broader implications beyond static classification; the features prioritized by the model could inform the design of more resilient nanowire architectures or eventually guide adaptive data acquisition. Ultimately, MEDA establishes a robust, measurement-efficient foundation for translating theoretical topological invariants into practical experimental diagnostics.

\section*{Acknowledgments}
This work is supported in part by the South Carolina Quantum Association, U.S. National Science Foundation under Award No. 2518605, the Clemson Palmetto 2 Cluster, and Clemson Creative Inquiry.

The authors used Google Gemini, Claude.ai and ChatGPT for language editing in all sections of the article. All content was reviewed and edited by the authors, who take full responsibility for the final work. 

\bibliographystyle{IEEEtran}
\bibliography{references}

\end{document}